\documentclass[12pt]{article}
\usepackage{array,multicol}
\usepackage{afterpage,float,flafter}
\usepackage[dvips]{epsfig,rotating}
\usepackage{pifont,fancybox}
\def\CellGroup{\bgroup}
\def\endCellGroup{\egroup}

\newcommand{\be}{\begin{equation}}
\newcommand{\ee}{\end{equation}}

\newcommand{\bea}{\begin{eqnarray}}
\newcommand{\eea}{\end{eqnarray}}
\newcommand{\eq}[1]{eq.~(\ref{#1})}

\newcommand{\gsim}{\ \rlap{\raise 2pt\hbox{$>$}}{\lower 2pt \hbox{$\sim$}}\ }
\newcommand{\lsim}{\ \rlap{\raise 2pt\hbox{$<$}}{\lower 2pt \hbox{$\sim$}}\ }

\newcommand{\matr}{\left( \begin{array}}
\newcommand{\ematr}{\end{array} \right)}

\newcommand{\np}[1]{Nucl. Phys. #1}

\def\beq{\begin{equation}}
\def\eeq{\end{equation}}
\def\bea{\begin{eqnarray}}
\def\eea{\end{eqnarray}}
\def\bq{\begin{quote}}
\def\eq{\end{quote}}
\def\ben{\begin{enumerate}}
\def\een{\end{enumerate}}

\def\ie{{\it i.e.}}
\def\eg{{\it e.g.}}

\def\sin{{\rm sin}}
\def\cos{{\rm cos}}
\def\tan{{\rm tan}}

\def\Dslash{\not{\hbox{\kern-4pt $D$}}}
\def\dslash{\not{\hbox{\kern-2pt $\del$}}}

\def\cp1sub{\setlength{\unitlength}{8pt}\begin{picture}(2,1)\mbox{\scriptsize CP} \end{picture}}                                 % CP-conserving

\makeatletter
\@addtoreset{equation}{section}
\makeatother

\title{
\vspace*{-2.0cm}
\begin{flushright}
\normalsize{
FERMILAB-Pub-02/001-T \\
hep-ph/0201xxx
}
\end{flushright}
\vspace*{1.0cm}
{Neutrinos that violate $CPT$,\\
and the experiments that love them}
\vspace*{0.8cm}
\author{\large\textbf
{G.~Barenboim$^a$, L.~Borissov$^b$, J.~Lykken$^{a,c}$}\\ 
\\
$^a$\normalsize\emph{Fermi National Accelerator Laboratory,
P.O. Box 500, Batavia, IL 60510, USA }\\
$^b$\normalsize\emph{Columbia University, New York, NY, 10027, USA}\\
$^c$\normalsize\emph{Enrico Fermi Institute, Univ. of Chicago, 5640
S. Ellis Ave., Chicago, IL 60637, USA}\\ }
}

\begin{document}
\maketitle

\vspace*{2cm}

\begin{abstract}
Recently we proposed a framework for explaining the observed evidence for
neutrino oscillations without enlarging the neutrino sector, by
introducing $CPT$ violating Dirac masses for the neutrinos.
In this paper we 
continue the exploration of the phenomenology of
$CPT$ violation in the neutrino sector.
We show that our $CPT$
violating model fits the existing SuperKamiokande data at
least as well as the standard atmospheric neutrino oscillation
models. We discuss the challenge of measuring $CP$ violation
in a neutrino sector that also violates $CPT$. We point out that
the proposed off-axis extension of MINOS looks especially
promising in this regard. Finally, we describe a method
to compute $CPT$ violating neutrino effects by mocking them up with
analog matter effects.
\end{abstract}

\thispagestyle{empty}
\newpage

\section{Introduction} 
As discussed in \cite{we} (see also \cite{Murayama}),
$CPT$ violation has the potential to 
explain all existing neutrino anomalies without either enlarging the
neutrino sector or introducing other new degrees of freedom.
The beauty and economy of this framework cannot escape
the reader who recalls that sterile neutrinos
where introduced into this game with the unique purpose of
explaining all observed anomalies with oscillations. 
Furthermore, if $CPT$ is violated by non-Standard Model dynamics, 
neutrinos are the most natural messengers of this breaking,
which does not require a concomitant breaking of Lorentz
invariance.

As in any model designed to include an explanation of the 
appearance signal observed in LSND, the most sonorous confirmation
of our proposal will arise with the confirmation of LSND 
itself by MiniBooNE \cite{miniboone}. 
While this will not be enough in itself to claim
that $CPT$ is violated, the smoking gun of our model
is that MiniBooNE will
see an appearance signal only when running in the antineutrino
mode and not in the neutrino one.  

\begin{figure}[t]
\vspace{-2.5cm}
\centerline{\epsfxsize 14.2cm \epsffile{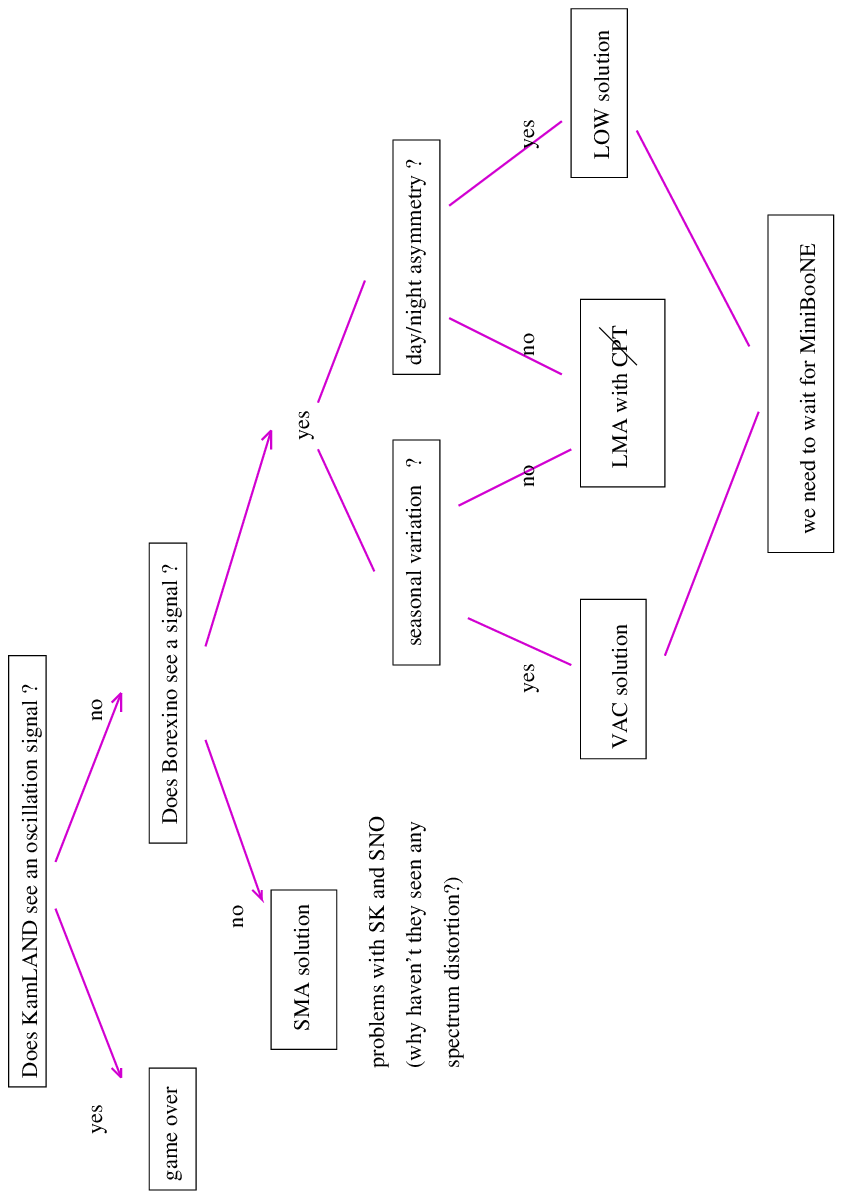}}
\caption{Flow chart for discovering $CPT$ violation by combining
the results of KamLAND and Borexino}
\label{chart}
\end{figure}

However, as we outlined in \cite{we}, this is by no means the unique
way to get evidence of $CPT$ violation in the neutrino sector.
We can take a shortcut
to the $CPT$ violating path by combining the information of
KamLAND \cite{kamland}
and Borexino \cite{borexino} (see fig \ref{chart}).

The Kamioka mine is now the home of KamLAND,  an experiment whose
principal goal is to confirm and pin down the mass difference
involved in the solar neutrino oscillations (provided this 
mass difference lies in the large mixing angle (LMA) region), 
by studying the flux and energy spectra of neutrinos produced by 
Japanese commercial nuclear reactors. As the best fit point
to all the neutrino experiments lies precisely in this region,
there is a growing consensus that the LMA zone is definitely the 
right place to look. However, if $CPT$ is violated, KamLAND might
be exploring the right place (LMA solution for neutrinos) with the 
wrong tool (reactor neutrinos, \ie\  , antineutrinos). 
According to our model, KamLAND will not see an oscillation signal, 
even if the mass difference involved in the solar neutrino 
oscillations lives in the LMA region.
However, this evidence by itself will not be hailed as evidence of $CPT$
violation. It will just drive the $CPT$ conserving believers 
to regions in the parameter space that do not receive today
the favor of the public, such as the LOW solution \cite{andre}.

A confirmation of the fall of the last discrete symmetry
might come nevertheless while combining this information with
data from the Borexino experiment.
Borexino is a solar neutrino real-time experiment at LNGS 
(Laboratori Nazionali del Gran Sasso) that makes use of the 
neutrino-electron scattering
reaction to detect neutrinos emitted from the Sun.
From the point of view of our $CPT$ violating model, Borexino
will explore the right place with the right tool.  
The experiment is mainly interested in the observation of the 
higher energy 7Be neutrinos, which produce a monochromatic line at 
863 keV. This line is predicted by all the standard solar models 
to be the second most important neutrino production
reaction (after the basic $pp$ reaction) in the Sun. The flux of 7Be 
neutrinos is predicted much more accurately (with an uncertainty 
of less than 10\%) and is about a 1000 times larger than the 8B neutrino
flux that is measured by SuperKamiokande and SNO. Also, since
7Be decay produces only neutrino lines, 
the theoretical predictions of neutrino oscillations are more unique 
for 7Be than for the 8B neutrinos, which have a broad energy spectrum 
(0-15 MeV). 

Borexino will see a signal inconsistent with background
only if the solar neutrino solution involves a large mixing
angle, \ie , one of LMA, LOW, vacuum oscillations (VAC) or
quasi-vacuum oscillations (QVO);
for the small mixing angle (SMA) solution, the neutrino
rate at Borexino will be suppressed almost down to the background
level. Given a signal, Borexino can distinguish between
different large mixing scenarios by looking at time variations,
in particular seasonal and diurnal variations.
The distinctive feature of a LOW solution 
will be earth matter effects which give diurnal variations, while
the QVO and VAC regions both offer seasonal variations. Therefore,
if Borexino does see a signal, and does not see either seasonal
variation or day/night asymmetry, while KamLAND sees an oscillation
signal,
this will undoubtedly point
towards a $CPT$ violating spectrum with an LMA solution for the
solar neutrinos.  On the other hand, if either a seasonal 
or diurnal variation is observed at Borexino, 
we should wait till 
MiniBooNE closes the discussion about $CPT$ in the LSND
region (one way or the other).

At this point a word of caution is in order, as there exists a very
small region (disfavored by the state of the art fits) 
where the LOW solution becomes the QVO one and in which no unmistakable
signal can be observed. To completely rule out this particular
point (which has a very low goodness of fit), the full capability of
the near future experiments must be used, \ie\   a day/night effect
will be detected by
KamLAND, after KamLAND is converted into a solar neutrino experiment.

%\begin{figure}[!ht]
%  \begin{center}
%\hspace{-5cm}{
%  \epsfig{file=chart.eps,width=16cm,angle=270}}
%\caption{ }
%\label{chart}
%  \end{center}
%\end{figure}

\section{Atmospheric vs (anti) atmospheric}

Since SuperKamiokande (SK) is a water Cerenkov experiment it simply adds up
all the neutrino and antineutrino information without distinction.
One wonders then if there is any possibility of digging out
from their data any hint about or constraint on the $CPT$ violation
in the atmospheric sector. With this goal in mind, we have performed
a selective $\chi^2 $ fit to SK multi GeV and sub GeV data
(a total of 40 data points),
where  
\be
  \chi^2_{\rm atm}= \sum_{M, S}\sum_{\alpha=e,\mu}\sum_{i=1}^{10} 
     \frac{(R_{\alpha,i}^{\rm exp}-
     R_{\alpha,i}^{\rm th})^2}{\sigma_{\alpha i}^2} \quad .
\ee
Here $\sigma_{\alpha,i}$ are the statistical errors,
the ratios $R_{\alpha,i}$ between the observed and predicted signal 
can be written as
\be
  R_{\alpha,i}^{\rm exp}= N_{\alpha,i}^{\rm exp}/N_{\alpha,i}^{\rm MC}
\ee
(with $\alpha$ indicating the lepton flavor and $i$ counting the
different bins, ten in total)
and $M,S$ stand for the multi-GeV and sub-GeV data respectively. 
As we have closely followed the spirit of the calculation in \cite{amol},
we refer the reader to this article for details and skip the technicalities. 

Since the parameter space is so huge (two mass differences, three mixing
angles and one $CP$ violating phase in each sector), we decided 
to make some simplifying assumptions 
which we believe will not have any impact 
on the results.
\begin{itemize}
\item All the $CP$ violating phases have been set to zero.
\item The mass difference related to LSND (the largest mass difference
in the antineutrino sector) is fixed to some arbitrary value.
For the energies and distances involved in the atmospheric neutrino
experiment this mass difference corresponds to a rapid oscillation,
and therefore
its exact value is not relevant provided it is large enough.
\item The mass difference involved in solar neutrino oscillations
(the smaller in the neutrino sector) is fixed to its best-fit point in 
the LMA region, \ie\  , $s_{12}^2=.29$ and $\Delta m_{\odot}^2= 4.5 \times
10^{-5} \mbox{ eV}^2$. 
\end{itemize} 
We are left therefore with seven parameters to fit: the 
neutrino and antineutrino mass differences giving the leading
contribution to the atmospheric oscillations, $\Delta m_{\rm atm}^2$ and
 $\Delta \bar{m}_{\rm atm}^2 $ respectively,
the corresponding mixing angles,
the connecting angle in the neutrino sector $s_{13}$, and the remaining
two angles in the antineutrino sector $\bar{s}_{23}$ and  
$\bar{s}_{13}$. For the sake of clarity Figure 2 provides a 
dictionary to our way of labeling the masses. As we label the masses
from the bottom to the top, the lightest state
always being
$m_1$ (or $\bar{m}_1$),  and the heaviest $m_3$ (or $\bar{m}_3$),
the mass difference involved in atmospheric
oscillations is $\Delta m_{23}^2$ in the neutrino case (with
mixing angle $\theta_{23}$), while the one in the antineutrino 
channel is $\Delta \bar{m}_{12}^2$ (with an effective mixing
angle $\sin(2 \; \bar{\theta}_{\rm atm}) \simeq 4 \bar{U}_{\mu 1}^2 
\bar{U}_{\mu 2}^2 \;$). 
Remember that in the neutrino case
$\Delta m_{12}^2$ and $\theta_{12}$ drive the solar neutrino oscillations,
while $\Delta \bar{m}_{23}^2$ and  $\sin(2 \; \theta_{LSND}) \simeq 4 
\bar{U}_{e 3}^2 \bar{U}_{\mu 3}^2 $ will account
for the LSND signal.

\begin{figure}[ht]
\vspace{1.0cm}
\centering
\epsfig{file=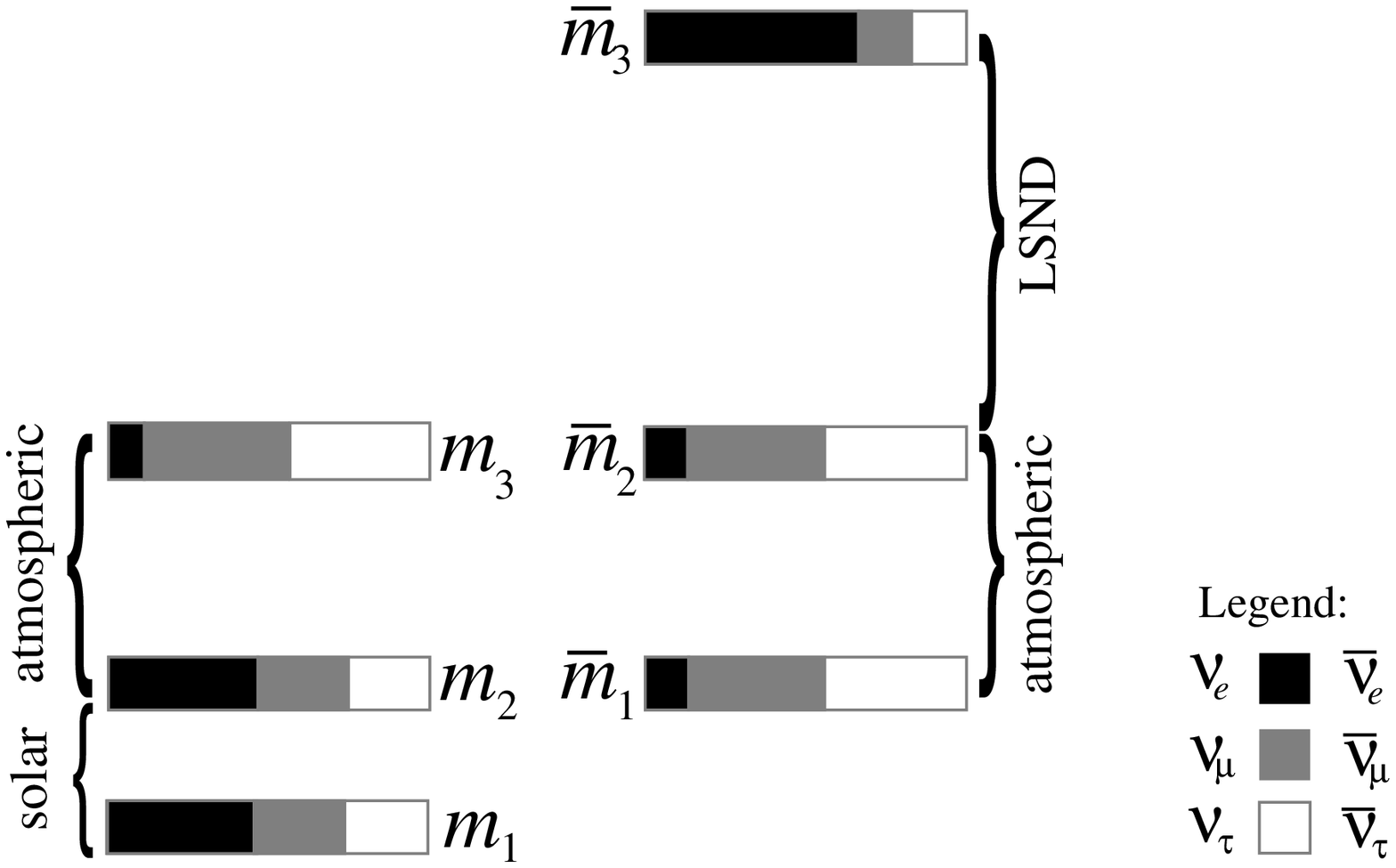,width=10cm}
\caption{ Typical $CPT$ violating (hierarchical) spectrum, able to 
account for the LSND, atmospheric and solar neutrino evidence }
\label{neutrino-spectrum}
\end{figure}

Let us remind the reader that the 
$\bar{s}_{13}$ angle is constrained by the CHOOZ experiment to be 
either close to zero or to one, as
\bea
     P_{\rm CHOOZ} & = & 1-4\bar{U}_{e3}^2 (1-\bar{U}_{e3}^2) 
      \sin^2 \left( \frac{\Delta m_{LSND}^2L}{4E} \right) 
      -4 \bar{U}_{e1}^2 \bar{U}_{e2}^2 \sin^2 
\left( \frac{\Delta \bar{m}_{\rm atm}^2 L}{4E} \right) 
	\nonumber \\ 
       & \simeq & 1-2 \bar{s}_{13}^2 \bar{c}_{13}^2 \; ,
\eea 
and in order to explain the LSND signal the latter solution will be needed.
Summing up, we have seven parameters and 40 data points for which
a scan over 100,000 points has found four regions with
comparable goodness of fit and a $\chi^2$ of about 39. One is centered
at
\bea
s_{23}^2= .40 \;\;\;\;\; , \;\;\;\;\; s_{13}^2= .01  \;\;\;\;\; , \;\;\;\;\;
\Delta m_{\rm atm}^2= 4 \cdot 10^{-3} {\rm eV}^2 
\eea
for the neutrino spectrum and
\bea
\bar{s}_{12}^2 = .74  \;\;\;\;\; , \;\;\;\;\; \bar{s}_{23}^2 =.98 
\;\;\;\;\; , \;\;\;\; 
\bar{s}_{13}^2 = .90  \;\;\;\;\; , \;\;\;\; \Delta \bar{m}_{\rm atm}^2=
4\cdot 10^{-3} {\rm eV}^2 \nonumber
\eea
for the antineutrino spectrum, while the other three live around
\bea
s_{23}^2= .40 \;\;\;\;\; , \;\;\;\;\; s_{13}^2= .01  \;\;\;\;\; , \;\;\;\;\;
\Delta m_{\rm atm}^2= (2.9, 2.6 \;{\rm and}\; 2.3) \cdot 10^{-3} {\rm eV}^2 
\eea
for the neutrino spectrum and
\bea
\bar{s}_{12}^2 = .74  \;\;\;\;\; , \;\;\;\;\; \bar{s}_{23}^2 =.98 
\;\;\;\;\; , \;\;\;\; 
\bar{s}_{13}^2 = .90  \;\;\;\;\; , \;\;\;\; \Delta \bar{m}_{\rm atm}^2=
(5.8, 6.6\;{\rm and}\; 7.6) \cdot 10^{-3} {\rm eV}^2 \nonumber
\eea
for the antineutrino spectrum respectively,
which implies a $\chi^2/{\rm d.o.f} \simeq 1.2 $.
This should be compared
with the result obtained (using the same program) for the CP
conserving case of $\chi^2 = 48 $ with 
$\chi^2/{\rm d.o.f} \simeq 1.3 $, where now
the number of degrees of freedom is not 33 as before but 37, as
only three parameters entered into the fit. In order to make
a fair comparison we have fixed in both cases the solar parameters.

\begin{figure}[htb]
\vspace{1.0cm}
\centering
\epsfig{file=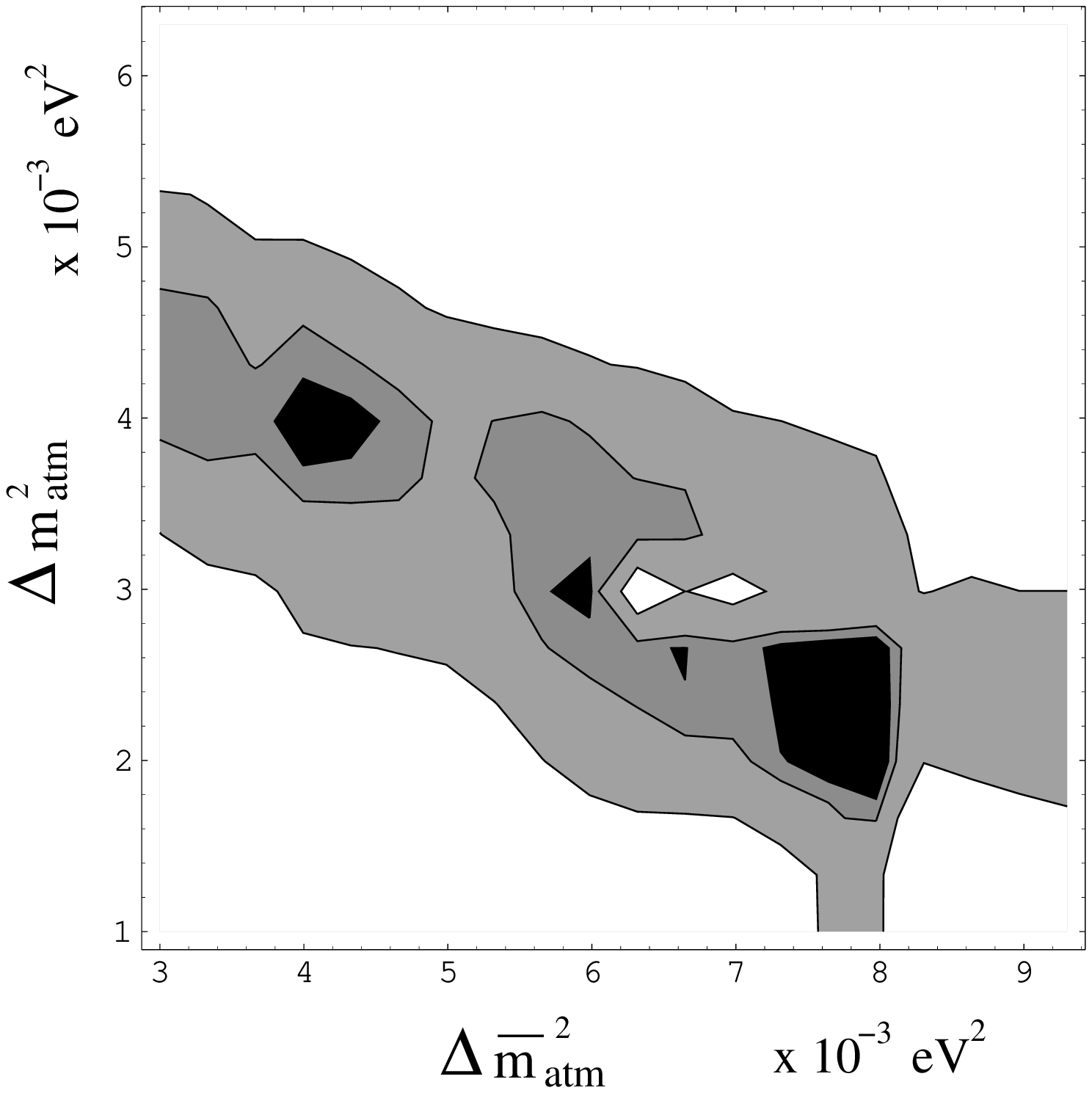,width=10cm,angle=270}
\caption{Best-fit regions for the SK atmospheric neutrino data in
the $\Delta \bar{m}_{\rm atm}^2$ - $\Delta m_{\rm atm}^2$ 
plane. The allowed regions are shown at 90\%, 95 \% and 99\% CL
with respect to the global minimum.}
\label{SK-fit}
\end{figure}

The ``better'' fit (in terms of a lower $\chi^2$ ) 
of the $CPT$ violating case is not surprising as more degrees of
freedom are available and therefore better agreement with the data
can be expected. However as both goodness of fit are similar
the most we can say is that both schemes are equally good, at least
from the SK point of view.  As expected SK data are a better 
constraint for neutrinos than for antineutrinos as the combination
of a lower cross section and a lower flux make the oscillation signal a
predominantly neutrino one.  Notwithstanding the above, a
correlation between the mass differences was found 
as can be seen in Figure 3. For one of the best-fit regions
both mass differences are almost equal, while for the other regions the
neutrino mass difference is almost half (or one third of) the 
antineutrino one; the neutrino mass difference
in this case coincides with the SK $CPT$ conserving best-fit point.
In all of our best-fit regions there is a large $CPT$ violating
difference in the mixing angles. 

One should notice however that our best-fits points are wildly
disfavored by CHOOZ, whose results were not taken into account in the 
fit. If one now imposes the CHOOZ bound, the best-fit regions
remain approximately the same, but the  $\chi^2$ grows to 
values around 44. Nevertheless, the goodness of the fit,
taken as $\chi^2/{\rm d.o.f}$ becomes now approximately 1.3,
and therefore is still as good as  the CPT conserving case.

Our fit confirms the expectations of \cite{weiler}, but
appears rather different
from the findings of Ref~\cite{solveig}.
Some difference from Ref~\cite{solveig} is expected since
that work used a
two generation approximation and didn't include matter effects.
More interesting, Ref~\cite{solveig}
allows the overall $\nu_{\mu}/\nu_e$ flux ratio to vary freely, 
a fact that has already proven to have a strong impact on the results.
Specifically, in a $CPT$ violating scenario varying this
parameter pushes the fit to large values of 
$\Delta \bar{m}_{\rm atm}^2$ ($\gsim 0.1 eV^2$), where the
rapidly oscillating antineutrino contribution washes out,
becoming essentially equivalent to a shift in the flux ratio.
This possibility is best regarded as complementary to our
results.

As a closing remark we would like to emphasize that ours
was a coarse grain fit and would need improvement to compare with
the state of the art for such analyses. The grid resolution on 
Figure 3 is $ 0.33 \times 10^{-3} \; eV^2$ with respect to 
$\Delta \bar{m}_{\rm atm}^2$ and $\Delta m_{\rm atm}^2$.
While the shape and possibly even the number of minima may change with a
finer resolution scan, we expect the overall correlation between the
parameters evident from the figure to remain.

\section{$CP$ vs $CPT$}

In a  picture containing three oscillating Dirac neutrinos,
a neutrino state of definite flavor $\alpha$, owner of
well defined weak interaction properties, is related to 
neutrino states of definite mass $m_k$ by
\begin{equation}
\nu_\alpha = \sum_k U_{\alpha k} \nu_k
\label{uno}
\end{equation}
\noindent
where $U$ is the unitary mixing matrix which, for 3 families, 
depends on 3 mixing angles and  1 $CP$ phase. It is clear that 
in the $CPT$ conserving case, the mixing matrices in the
neutrino and antineutrino sector are not independent, since
one is the conjugated of the other. However, if $CPT$ is no
longer a good symmetry, both matrices are not related and
an incredible rich experimental potential emerges. Let's
then follow the game for awhile to see
what are the smoking guns we are looking for.

If the ``$\alpha$" state is born  at $t = 0$, the probability 
amplitude that, at time $t$, it will end up  as the ``$\beta$" 
state is
\begin{equation}
A (\alpha \rightarrow \beta ; t) = \sum_k U_{\alpha k} U_{\beta k}^*
exp[- i E_k t] \quad .
\label{dos}
\end{equation}
It can be seen  from (\ref{dos}) that the time-dependent amplitude 
contains the interference of different ``$k$" terms, with different
 weak phases in $U_{\alpha k} \, U_{\beta k}^*$ and different 
oscillation phases governed by $E_k$, precisely 
the necessary ingredients to generate $CP$ violation
in the oscillation probability. If we now impose $CPT$, 
the amplitude for conjugated flavor states satisfy 
\begin{equation}
A (\bar{\alpha} \rightarrow \bar{\beta}; t) = \sum_k U_{\alpha k}^*
U_{\beta k} \, exp[-i E_k t]
\end{equation}
so that, $CPT$ implies 
\begin{equation}
A (\bar{\alpha} \rightarrow \bar{\beta}; t)
= A^* (\alpha \rightarrow \beta; - t) 
\end{equation}  
On the other hand, the $CP$ transformation relates 
the probabilities for the original transition and its conjugate,
\begin{equation}
|A (\alpha \rightarrow \beta; t)|^2 =
|A (\bar{\alpha} \rightarrow \bar{\beta}; t)|^2
\end{equation}
while the T invariance relates the probabilities for the
original transition and its inverse 
\begin{equation}
\begin{array}{ll}
& | A (\alpha \rightarrow \beta ; t)|^2 = | A (\beta
\rightarrow \alpha; t)|^2\\
& |A (\bar{\alpha} \rightarrow \bar{\beta}; t)|^2 =
|A (\bar{\beta} \rightarrow \bar{\alpha}; t)|^2
\end{array}
\end{equation} 
Therefore, in a $CPT$ conserving world, $CP$ and $T$ violation effects
can take place in appearance
experiments only. For disappearance experiments, $\beta = \alpha$,
and Eq.~(\ref{dos}) implies
\begin{equation}
A^* (\alpha \rightarrow \alpha ; t) = A (\alpha \rightarrow \alpha;
- t)
\end{equation}
As a consequence, no $CP$ or $T$ violation effect can be manifested in reactor
or solar neutrino experiments (in a $CPT$ conserving scenario).

In a $CPT$ violating scenario however, even the survival probabilities
for the conjugated channels can be different, opening the door to
a new world of measurements but closing the path to the possibility
of measuring $CP$ violation using conjugated channels. So far, most
(if not all) the proposals of measuring the $CP$ violating phase
rely precisely on this technique, \ie\    first assume $CPT$ and
then construct an asymmetry with the different channels.
Therefore, if the physics which hides beyond the Standard Model
does not conserve $CPT$, these asymmetries will confirm
that neutrinos are antineutrinos have an independent spectra
but will not provide any single clue to whether $CP$ is violated.
The question will be then, whether there is an experiment (besides
the ones described in the previous section) able to not only test
$CPT$ by itself (not mixing results from different experiments) but
also to measure genuine $CP$ violation.

The answer is yes, such an experiment exists: it is MINOS \cite{minos}.
It will search for neutrino oscillations and measure with unprecedent
precision the muon neutrino survival probability. MINOS
can run in neutrino and antineutrino modes; it will measure both
survival probabilities and pin down the mass difference involved
in the atmospheric neutrino signal in both channels independently
with great accuracy, thus providing a self-consistent test
of $CPT$. One should bear in mind that due to the difference
in cross section and production, one ends up with approximately
six more times neutrino than antineutrino signal, and therefore any
$CPT$ comparison must involve a sizeable amount of running time 
in the antineutrino mode. However this is independent of whether
$CPT$ is conserved or not, and has been already taken into account 
when planning to measure the $CP$ phase by combining results from
conjugated channels.

On top of that, the recent development of the off-axis beam 
ideas \cite{off} 
(neutrinos emitted at angles 10-20 mrad with respect to the
beam axis create an intense beam with well defined energy) provide
the possibility of $CP$ violating studies without resorting to
conjugated channels. In this case the idea will be to measure
in two detectors (Soudan mine and Lake Superior)  the electron
neutrino appearance probability and (with some knowledge of
the connecting angle $s_{13}$ or by just measuring the
two values for the transition probability) extract the value of 
the $CP$ phase. Remember that this angle is not constrained by 
reactor experiments (\eg\ CHOOZ) as these experiments involve 
antineutrinos and therefore can be sizeable. The two detector proposal
has the advantage that is not based on the assumption that
$CPT$ is conserved, and that (as it does not involve antineutrinos)
more precision can be reached with less running time.

Apart from the man-made neutrinos, one can use directly
the nature-made atmospheric neutrinos to check the status
of the $CPT$ symmetry. This is precisely the idea behind
MONOLITH \cite{monolith}. 
This experiment will compare the event rates induced by 
the near and far (downward and upward) atmospheric muon neutrino
fluxes (exactly as SuperKamiokande does) but with an iron
detector, and therefore
will be able to constraint the neutrino and antineutrino mass
differences independently. 

All in all, although it is certainly against our prejudices, it is
not entirely an anathema to propose that a breakdown of $CPT$ 
invariance might be responsible for all the experimental evidence
in the neutrino sector that has been found so far and cannot
be explained within the Standard Model (or even its
minimal extensions). Within this scheme, the mixing matrices
$U$ and $\bar{U}$ are unitary but not related to each other.
The Kobayashi-Maskawa argument on the ambiguity in the phases of 
fermion fields reduces the number of independent real parameters
to four each (the mixing angles plus one phase) for 
$U$ and $\bar{U}$ . Thus altogether there are 14 real parameters 
describing neutrino oscillations, three masses (two mass differences) 
and four mixing parameters for neutrinos and likewise for antineutrinos
all of which can be  determined in the forthcoming experiments. 
It is true that the obstacles in this task are high, but it is
also true that the possible insight gained is even higher!

\section{Resuscitating the ether}

Although the $CPT$ violating idea is tempting,
it suffers from the drawback of being impractical for calculations.
As any local Lorentz invariant field theory
automatically conserves $CPT$, in order to discuss
$CPT$ violation, we must move to an operator Hamiltonian description
in momentum space (as shown in ~\cite{we}). Therefore
by adopting $CPT$ violation we have lost more than fifty years of
developments in quantum field theory, and are back to square
one for any calculational purpose.  One might wonder then whether
there is any possibility of keeping the utility of local field theory
but not its restrictions, \ie\   to have an effective field theory that
mimics in some way $CPT$ violation.  In fact, this possibility 
does exist, and it has been known for quite a long time. 
Matter effects are the key to an effective 
field theory description of $CPT$ violation.
                              
When neutrinos propagate through matter, the forward scattering of 
neutrinos off the background matter will induce an index of refraction 
for the neutrinos (which is different from that of antineutrinos). 
The neutrino index of refraction will depend generally on the flavor
(electron, muon and tau neutrinos will have different indices of refraction
because the background matter contains different amounts of each of them).
The index of refraction acts like an effective mass
term. Thus the effects of $CPT$ violation, for the purposes of calculation,
can be modeled by a kind of ``ether'' populated by different concentrations
of ether-matter, giving different effective masses for
neutrinos and antineutrinos.
For the sake of clarity, in the following we will illustrate our
point using the passage of neutrinos through standard matter,
\ie\   matter composed only by electrons (no muons or taus). However the
extension to matter containing heavy leptons is straightforward.

Background electrons in normal matter will interact via charged
currents with electron neutrinos as
\bea
{\cal L}_{\rm cc} = \frac{G_F}{\sqrt{2}} \bar{e} \gamma^\mu 
\left( 1 - \gamma_5 \right) \nu_e \;\; \bar{\nu}_e \gamma_\mu  
\left( 1 - \gamma_5 \right) e \; ,
\eea
which after a Fierz rearranging looks like
 \bea
{\cal L}_{\rm cc} = \frac{G_F}{\sqrt{2}} \bar{\nu}_e \gamma^\mu 
\left( 1 - \gamma_5 \right) \nu_e \;\; \bar{e} \gamma_\mu  
\left( 1 - \gamma_5 \right) e \;.
\eea
For a medium with electrons at rest, we have
\bea
\bar{e} \gamma_\mu  \left( 1 - \gamma_5 \right) e = \delta_{\mu 0} N_e
\eea
where $N_e$ is the electron number density. This interaction is
equivalent to a repulsive potential, $V= \sqrt{2} G_F N_e$, for
left-handed neutrinos given by the wave function $\frac{1}{2} 
\left( 1 - \gamma_5 \right) \nu_e$. For relativistic neutrinos,
with $E \simeq p + m^2/2p$, this potential amounts to an effective
mass for the electron neutrino 
\bea
m^2_{\rm eff} \equiv A = 2 \sqrt{2} G_F N_e E.
\eea
In the flavor basis the potential is diagonal. Restricting ourselves
to the two generation case for simplicity, the mass matrix becomes
\bea
M^2 =\left( U \pmatrix{ m_1^2 & 0 \cr 0 & m_2 ^2 \cr} U^\dagger +
\pmatrix{ A & 0 \cr 0 & 0 \cr} \right)
\eea
where $U$ is the unitary transformation between the flavor
and mass bases and $A$ acts like an induced mass (squared)
for the electron neutrino from the propagation through a background
of electrons. The corresponding expressions in the antineutrino case
can be obtained by the replacements, $ A \longrightarrow -A $ and
$U \longrightarrow U^*$.

$M^2$ can be diagonalized  by $U_m$, the mixing matrix in 
the medium:
\bea
U^\dagger_m M^2 U_m \equiv \pmatrix{ M_1^2 & 0 \cr 0 & M_2^2 \cr}
\eea
where
\bea
M_{2,1}^2 = \frac{ (\Sigma + A ) \pm \left[ (A - \Delta \cos(2 \theta))^2 +
(\Delta \sin(2 \theta) )^2 \right] }{2}
\eea
with $\Sigma = m_2^2 + m_1^2 $, $\Delta = m_2^2 - m_1^2 $, $\theta$ 
is the mixing angle in vacuum and 
\bea
U_m = \pmatrix{ \cos\theta_m & - \sin\theta_m \cr
\sin\theta_m & \cos\theta_m \cr}
\eea
where $\theta_m$ is the mixing angle in matter and is given by
\bea
\tan 2 \theta_m = \frac{ \Delta \sin(2 \theta)}{-A + \Delta \cos(2 \theta)}\; .
\eea
To make our point even more transparent, 
let us assume that $ m_1 \simeq m_2 = m $.
The neutrino masses in vacuum become
\bea
M_1^2= m^2 + A  \;\;\;\;\;\; M_2^2 = m^2
\eea
while the antineutrino masses are given by
\bea
\bar{M}_1^2= m^2 - A    \;\;\;\;\;\; \bar{M}_2^2 = m^2 \; .
\eea
Even in this extremely simplified model a
drastic breakdown of $CPT$ can be obtained. The complete scheme,
including the three families and effective densities for
all the charged leptons, can accommodate a large
subclass of $CPT$ violating spectra
(although no analytic formulae can be expected in this case).

It is clear then that in a typical medium such as the Earth or the Sun,
neutrinos and antineutrinos have $CPT$ violating spectra. Therefore,
in order to describe the $CPT$ violating 
extension of the Standard Model, and thus to account for all the existing
neutrino anomalies which we have presented in \cite{we}, we have
to only choose the electron (muon and tau) density that is appropriate 
to describe the $CPT$ violating 
mass difference that we need. Calculation of physicial processes that
we are interested in can then be performed with standard
field theory techniques.  

\section{Conclusions} 

$CPT$ violation has the potential to explain all the existing
evidence about neutrinos with oscillations to active flavors. 
Such a scenario makes specific (and unique) predictions that 
will be tested in the present round of neutrino experiments.
$CPT$ violation can be searched for independently of whether it occurs
in conjunction with $CP$ violation or not. As we have shown,
both symmetries can (and must) be tested separately. 
So far, we have no evidence of $CPT$ conservation in the neutrino
sector. Indeed as we have shown all the existing data,
including from SuperKamiokande, is most economically explained
if $CPT$ is broken. The true status of $CPT$ in the neutrino
sector can be established by the
combined results of KamLAND and Borexino, or by MiniBooNE.
In the atmospheric sector MINOS is the ideal experiment for 
such a test. 

From a practical point of view, all the calculational inconveniences
of a $CPT$ violating model can be avoided (for a large
subclass of models)
by constructing an analog effective
theory of neutrinos propagating in a medium with a density
such as to reproduce the desired mass pattern.

Certainly, there are many exciting features and potential signatures for
models with $CPT$ violation. We leave it to
the reader to judge the
degree of skepticism that is appropriate when considering the phenomenology
of these theories, which disobey the eleventh commandment,
\ie\  thou shalt conserve $CPT$.

\subsection*{Acknowledgements}
\noindent
We are grateful to P. Mackenzie and J. Simone for sharing with us
their lattice expertise and giving us access to their lattice
cluster. We thank John Beacom, Nicole Bell,
Janet Conrad, Andre de Gouvea, Boris Kayser 
and Solveig Skadhauge for comments and suggestions.
Research by GB and JL was supported by the U.S.~Department of Energy
Grant DE-AC02-76CHO3000 while that of LB by the Sloan Foundation.


\begin{thebibliography}{99}
 
\bibitem{we}
G.~Barenboim, L.~Borissov, J.~Lykken and A.~Y.~Smirnov,
``Neutrinos as the messengers of CPT violation,''
arXiv:hep-ph/0108199.

\bibitem{Murayama}
H.~Murayama and T.~Yanagida,
%``LSND, SN1987A, and CPT violation,''
Phys.\ Lett.\ B {\bf 520}, 263 (2001)
[arXiv:hep-ph/0010178].

\bibitem{miniboone}
MiniBooNE Collaboration, A. Bazarki, \np{B 91} (Proc. Suppl.) (2000), 210.

\bibitem{kamland}
KamLAND Collaboration, A.Piepke, \np{B 91} (Proc. Suppl.) (2000), 99.

\bibitem{borexino}
The Borexino web page: almine.mi.infn.it/html/borexino.html.
 
\bibitem{andre}
A.~de Gouvea and C.~Pena-Garay,
%``Solving the solar neutrino puzzle with KamLAND and solar data,''
Phys.\ Rev.\ D {\bf 64}, 113011 (2001)
[arXiv:hep-ph/0107186].

\bibitem{minos}
The MINOS Technical Design Report, Fermilab report NUMI-L-337.

\bibitem{monolith}
MONOLITH Collaboration, N.Y. Agafonova et al., ``MONOLITH: a Massive 
magentized iron detector for neutrino oscillation studies'', 
LNGS-P26-2000, LNGS-P26-00, CERN-SPSC-2000-031, CERN-SPSC-M-657.

\bibitem{off}
K.~T.~McDonald,
``An off-axis neutrino beam,''
arXiv:hep-ex/0111033;
A.~Para and M.~Szleper,
``Neutrino oscillations experiments using off-axis NuMI beam,''
arXiv:hep-ex/0110032.

\bibitem{amol}
G.~Barenboim, A.~Dighe and S.~Skadhauge,
``Combining LSND and atmospheric anomalies in a three-neutrino picture,''
arXiv:hep-ph/0106002.

\bibitem{weiler}
V.~D.~Barger, S.~Pakvasa, T.~J.~Weiler and K.~Whisnant,
%``CPT odd resonances in neutrino oscillations,''
Phys.\ Rev.\ Lett.\  {\bf 85}, 5055 (2000)
[arXiv:hep-ph/0005197].

\bibitem{solveig}
S.~Skadhauge,
``Probing CPT violation with atmospheric neutrinos,''
arXiv:hep-ph/0112189.

\end{thebibliography}
\end{document}